\documentclass[a4paper,11pt]{article}
\usepackage{jcappub}
\usepackage{epsfig}
\usepackage{graphics}

\newcommand{\kumiko}{\color{blue}}

\title{The fate of ultrahigh energy nuclei in the immediate environment of young fast-rotating pulsars}
\author[a]{Kumiko Kotera}
\author[b]{Elena Amato}
\author[b,c]{Pasquale Blasi}

\affiliation[a]{Institut d'Astrophysique de Paris
UMR7095 -- CNRS, Universit\'e Pierre \& Marie Curie,
98 bis boulevard Arago
F-75014 Paris, France.}
\affiliation[b]{INAF/Osservatorio Astrofisico di Arcetri, Largo E. Fermi 5, I-50125 Firenze, Italy.}
\affiliation[c]{Gran Sasso Science Institute (INFN), Viale F. Crispi 6, 60100 L'Aquila, Italy.} 

\emailAdd{kotera@iap.fr}

\abstract{
{Young, fast-rotating neutron stars are promising candidate sources for the production of ultrahigh energy cosmic rays (UHECRs). The interest in this model has recently been boosted by the latest chemical composition measurements of cosmic rays, that seem to show the presence of a heavy nuclear component at the highest energies. Neutrons stars, with their metal-rich surfaces, are potentially interesting sources of such nuclei, but some open issues remain: 1) is it possible to extract these nuclei from the star's surface? 2) Do the nuclei survive the severe conditions present in the magnetosphere of the neutron star? 3) What happens to the surviving nuclei once they enter the wind that is launched outside the light cylinder? 
In this paper we address these issues in a quantitative way, proving that for the most reasonable range of neutron star surface temperatures ($T<10^7\,$K), a large fraction of heavy nuclei survive photo-disintegration losses. These processes, together with curvature losses and acceleration in the star's electric potential, lead to injection of nuclei with a chemical composition that is mixed, even if only iron is extracted from the surface. We show that under certain conditions the chemical composition injected into the wind region is compatible with that required in previous work based on purely phenomenological arguments (typically $\sim 50\%$ protons, $\sim 30\%$ CNO and $\sim 20\%$ Fe), and provides a reasonable explanation of the mass abundance inferred from ultra high energy data.}
}


\arxivnumber{}

\date{\today} 

\begin{document}
\maketitle
\flushbottom

\section{Introduction}\label{section:intro}

Magnetized and fast-spinning neutron stars were introduced early on as potential candidate sources of ultrahigh energy cosmic rays (UHECRs) \cite{Venkatesan97,Blasi00,Arons03}. The model has been revived recently \cite{Murase09,K11, FKO12,FKO13}, mainly because of new measurements of the chemical composition. These sources present many advantages compared to the more classical scenarios invoking gamma-ray bursts or Active Galactic Nuclei. First, the energy budget supplied by their population is relatively comfortable, due to their large rotational energy reservoir ($E_{\rm rot}\sim 2\times10^{52}\,{\rm erg}\,I_{45}P_{-3}^{-2}$, with $I$ the star inertial momentum and $P$ its spin period\footnote{Here and in what follows, quantities are labelled $Q_x\equiv Q/10^x$ in cgs units unless specified otherwise.}, for an isolated new-born pulsar spinning close to the disruption limit), and their population density ($\dot{n}_{\rm s}\sim 3\times 10^{-4}\,{\rm Mpc}^{-3}\,{\rm yr}^{-1}$  \cite{Lorimer08}). The energy injected into UHECRs is of order ${\cal E}_{\rm UHECR} \sim 0.5\times 10^{45}\,{\rm erg}\,{\rm Mpc}^{-3}\,{\rm yr}^{-1}$~\cite{Katz09}, which implies that a fraction of order $10^{-4}$ of the neutron star population is required to achieve the UHECR flux level. In addition, the latest results reported by the Auger Observatory point towards a chemical composition of UHECRs that is not compatible with a light composition at the highest energies \cite{Auger_icrc13,Auger_compo_2014a,Auger_compo_2014b}. The Telescope Array results seem to show the same trend within systematics \cite{TAicrc11,Pierog13,TA_auger_icrc13}. Unlike AGNs and GRBs for which heavy nuclei production seems to be challenging~\cite{Lemoine02,Pruet02,Horiuchi12}, neutron stars, with their metal-rich surfaces, offer a favorable site to produce heavy nuclei.

One natural question is whether heavy nuclei stripped off the surface can survive the immediate environment of the star. This question was examined in Refs. \cite{Protheroe98,Bednarek97,Bednarek02} for the purpose of calculating the associated fluxes of neutrinos and gamma-rays, focusing on nuclei accelerated up to $10^{15-16}\,$eV in known local pulsars such as Crab and Vela. These authors showed that the thermal radiation field emanating from the neutron star surface was an issue for the survival of high energy nuclei for surface temperatures $T>10^7\,$K. They also examined the effect of the non-thermal soft photon background on $10^{15-16}\,$eV nuclei, in the specific case of particle acceleration in the pulsar outer gap, at the light cylinder. 

Farther out, for distances much larger than the light cylinder, and for energies $E\gtrsim 10^{18}\,$eV, it was speculated that the curvature radiation losses might be avoided \cite{Venkatesan97,Arons03,FKO12}, and that the effects of the radiation field due to the nebula may not be such as to inhibit the escape of accelerated iron nuclei at the highest energies \cite{FKO12, LKP13,FKMO14_paper}. More important are usually the hadronic interactions in the supernova ejecta: Refs.~\cite{FKO12} and \cite{FKO13} showed that even when these are dominant, escape of the highest energy particles is still allowed and a successful fit to the observed data can be obtained by tuning the initial composition (typically $\sim 50\%$ protons, $\sim 30\%$ CNO and $\sim 20\%$ Fe).

In this study, we re-examine the fate of heavy nuclei in the neutron-star immediate environment, before they are injected in the wind for acceleration to ultrahigh energies. One of our objectives is to estimate the chemical composition that is produced by photo-hadronic interactions and channelled into the wind, when pure iron is extracted from the surface of the star. 

To this purpose we calculate the photo-hadronic processes when iron nuclei are injected in the pulsar thermal radiation background. In agreement with Ref.~\cite{Protheroe98}, we find that for the most reasonable range of neutron star surface temperatures ($T<10^7\,$K), a large fraction of heavy nuclei survives the losses on the radiation environment of the star. Moreover, photo-disintegration leads to the production of secondary particles (intermediate nuclei and nucleons) that may help achieving a successful fit to the UHECR data, as required in Ref.~\cite{FKO13}. The nuclei become then part of the pulsar wind, which is where most of the rotational energy lost by the star is thought to end up. The mechanism through which this wind is accelerated to its terminal Lorentz factor is not clear, but, from extrapolation of what we infer in young (though not newborn) pulsars, we assume that most of the wind energy is in the form of particle kinetic energy, rather than Poynting flux. For reasonable values of the pulsar multiplicity, the wind Lorentz factor is still high enough so as to guarantee that the heaviest elements have ultrahigh energy. We concentrate here on pulsars with rotation periods at birth $P\lesssim 10\,$ms and mild dipolar magnetic fields ($B\sim 10^{12-13}\,$G), as this parameter space was found to be ideal for successful acceleration of UHECRs and their escape from the surrounding supernova ejecta~\cite{FKO12}. The existence of such pulsars is supported by pulsar population studies \cite{Faucher06}, and could be confirmed by the observation of X-rays and gamma-ray signatures from ultra-luminous supernovae \cite{KPO13,Metzger14, Murase14}. Some authors also suggest that the radio-emission from the Crab nebula imply that the Crab pulsar was born with a 5 ms period \cite{Atoyan99}.

The paper is organized as follows: in Section~\ref{section:extraction} we discuss the processes through which particles can be extracted and accelerated within the light cylinder of a pulsar. Taking into account the various radiation losses at play, we calculate the maximum Lorentz factor that particles can reach, mainly determined by the balance between acceleration in the electric field of the pulsar and curvature losses (this Lorentz factor is important when compared with the threshold of photo-hadronic interactions). In Section~\ref{section:photodis}, we detail our calculations of the production of secondary products of photo-disintegration, that determine the chemical composition of particle injected in the wind. Section~\ref{section:spectra} sketches the later acceleration of particles in the wind, and presents the spectra that will be injected in the nebula and/or the supernova ejecta region for further processing. We discuss our results and conclude in Section~\ref{section:discussion}.

\section{Ion extraction and acceleration within the light cylinder}\label{section:extraction}

Neutron-star surfaces are believed to be composed of anisotropic, tightly bound condensed matter. Calculations show that the outer layers should be composed mainly of long molecular chains with axes parallel to the magnetic field \cite{Ruderman72,Chen74}. The chains are thought to be composed of $^{56}$Fe ions forming a one-dimensional lattice with an outer sheath of electrons. The binding energy of the ions can be estimated as $\sim 14\,$keV and the lattice spacing as $l\sim 10^{-9}\,$cm \cite{Ruderman75}. Ions can thus be stripped off the surface provided that they experience a surface electric field of ${\cal E}_0 = {14\,{\rm keV}}/({Z e l}) \sim 5.4\times 10^{11} \,Z_{26}^{-1}l_{-9}^{-1}\,{\rm V} {\rm cm}^{-1}$.

However, ions can also be stripped off with milder electric fields if the surface is bombarded by particles (for instance leptons accelerated towards the star surface after the cascading processes in the magnetosphere are started), and/or if they can be boiled off the surface by stellar heat~\cite{Ruderman75,Arons79}.
 In the case of a pulsar with millisecond periods at birth, the electric field that can be provided at the surface of the star can be estimated as (see, e.g., \cite{Arons03})
 \begin{equation}
 {\cal E} = \frac{2\pi R_\star B}{Pc}\sim 6.3 \times 10^{14}\,B_{13}R_{\star,6}P_{-3}^{-1}\,{\rm V\, cm}^{-1}\ ,
 \end{equation}
where $B$ is the dipole magnetic field strength of the star, $R_\star$ its radius and $P$ its rotation period. 
In practice, this electric field does not persist on large distances, as it is expected to be readily screened out by pair cascading. 

However the total potential drop that is available, at least in principle, for particle acceleration in the magnetosphere is\footnote{Throughout this work, we have adopted the convention used in Ref.~\cite{Arons03} for the geometric pre-factors of the electromagnetic energy loss rate. See in particular its footnote 5.}: 
\begin{equation}
\Phi = \frac{2\pi^2BR_\star^3}{P^2c^2} \sim 6.6\times 10^{19}\,  B_{13}R_{\star,6}^3P_{-3}^{-2}\, {\rm V}\ ,
\end{equation}
which corresponds to a maximum achievable particle Lorentz factor:
\begin{eqnarray}\label{eq:Emax}
 \gamma_{\rm fpd}=\frac{Ze}{Am_pc^2}\,\Phi\,=7\times 10^{10}\,\,\frac{Z}{A}\,B_{13}\,P_{-3}^{-2}\ .
\end{eqnarray}
In reality, the maximum energy achievable by ions within the corotating region will be limited by losses, the most severe among which is curvature loss. Assuming that the total potential drop $\Phi$ is available for particle acceleration over a gap of length $\xi R_{\rm L}$ with $R_{\rm L}=c P/(2\pi)\sim 4.8\times 10^{6}\,P_{-3}\,$cm the light cylinder radius, the equation describing the particle energy evolution, subject to acceleration and curvature losses can be written as:
\begin{equation}
\frac{{\rm d}\gamma}{{\rm d} t} = \frac{Ze{\Phi}}{Am_{p}c^2} \frac{2 \pi}{\xi P}-\frac{8\pi^2}{3c P^2}\frac{Z^2e^2}{Am_{\rm p} c^2}\ \gamma^4\ .
\label{eq:dgdt}
\end{equation}
The maximum Lorentz factor achievable for radiation limited acceleration is then found equating the two terms in the {\it r.h.s.} of Eq.~\ref{eq:dgdt}, and turns out to be:
\begin{equation}
\label{eq:gamma_curv}
\gamma_{\rm curv} = \left( \frac{3\pi BR_\star^3}{2 Z e c P \xi}\right)^{1/4}\sim 1.1\times 10^8 Z_{26}^{-1/4}\xi^{-1/4}B_{13}^{1/4}P_{-3}^{-1/4}R_{\star,6}^{3/4}\ .
 \end{equation}
The actual maximum energy that particles can reach at any time within the corotating magnetosphere will be set by:
\begin{equation}
\label{eq:gmax}
\gamma_{\rm max}=\min \left(\gamma_{\rm fpd},\gamma_{\rm curv}\right)\ .
\end{equation}

One thing to notice in this expression is that $\gamma_{\rm curv}$ depends very weakly on the fraction of $R_{\rm L}$ over which the gap extends ($\xi$). On the other hand, the acceleration time to a given energy, $t_{\rm acc}(\gamma)$, has a much stronger dependence on the unknown $\xi$:
\begin{equation}
\label{eq:tacc}
t_{\rm acc}(\gamma)=\frac{A m_{\rm p} c^2 \gamma}{Z e \Phi}\frac {\xi P}{2 \pi}=5\times 10^{-9} s\ \frac{\gamma}{10^6}\ \frac{A_{56}}{Z_{26}}\ B_{13}^{-1}\ R_{\star,6}^{-3}\ P_{-3}^3\ \xi\ .
\end{equation}
On the other hand, while in general $\xi$ is rather uncertain, depending on pair creation in the polar cap, the constraint  $R_\star/R_{\rm L}<\xi<1$ is particularly tight for the fast-spinning pulsars we are concerned with: in this case $R_\star/R_{\rm L}\sim 0.2\, R_{\star,6}P_{\rm i,-3}^{-1}$ implies $\xi={\cal O}(1)$ and that the gap cannot be far from the surface of the star.

In this paper, we will work under the most classical assumption that particles are accelerated close to the stellar surface. For scenarios in which the acceleration happens at the light cylinder or further out, the thermal background could be too anisotropic and diluted to play a role. On the other hand, for the outer gap scenario \cite{Cheng86,Cheng86_2}, a non-thermal soft photon background could exist that could play a similar target role as the thermal background \cite{Bednarek97,Bednarek02}. Due to the even larger amount of uncertainties it entails, we will not consider this scenario further in the paper. 

One of the key points we want to make in this paper is to stress the crucial role of the Lorentz factor of the wind: while in the literature one can often find the argument that pulsars cannot be sources of CRs with Lorentz factor above $\gamma_{\rm max}$ (Eq.~\ref{eq:gmax}), because of the dramatic curvature losses inside the light cylinder, we argue that if $\gamma_{\rm max}<\gamma_{\rm w}$ (where $\gamma_{\rm w}$ is the Lorentz factor of the wind, that we express later in Eq.~\ref{eq:gwind}), once particles end up in the wind, they get advected with it at the Lorentz factor $\gamma_{\rm w}$, irrespective of the energy they reached in the magnetosphere. We will discuss the implications of this point in Section~\ref{section:spectra}.

Notice also that while photo-disintegration of nuclei may become very important inside the light cylinder, these interactions conserve the Lorentz factor of the nuclei involved. Photo-pion production of protons can in principle change the Lorentz factor of individual nucleons, but this process is very slow and its effect is found to be negligible.

\section{Photo-hadronic interactions on the thermal radiation of the star}\label{section:photodis}
Ultrahigh energy ions can experience photo-pion production or photo-disintegration in the thermal radiation fields generated by the star. As we will see in the following, the latter process is much more important than the former and leads to the production of progressively lighter nuclei, with interactions that change the number of nucleons while conserving the particle Lorentz factor. Before discussing in detail the consequences of this process on the composition of the UHECR flux from a newborn pulsar, we first recall briefly our current knowledge on the temperature of the pulsar surface at early stages, and estimate the key quantities involved in the photo-hadronic processes. 

\subsection{Photo-hadronic interaction timescales}

The thermal structure of the neutron star upper layers is not well understood, and constraints from observations are scarce (e.g.,~\cite{Pavlov02,Potekhin03,Potekhin14}). Neutron stars with magnetized envelopes made of accreted material have local surface temperatures of typically $T\sim10^{5.5-7}\,$K, and for magnetized objects, show a profile that can vary as a function of the polar angle. 

The thermal photon energy density, $U_{\rm th}$, can be expressed as a function of the neutron star surface temperature $T$ as $U_{\rm th}= a T^4$, $a$ being the radiation constant. This thermal component peaks at energy $\epsilon_{\gamma,{\rm th}}=k_{\rm B}T\sim 86\,{\rm eV}\,T_6$. 

One can model the photo-disintegration process for nuclei with a delta function (representing the Giant Dipole Resonance with a threshold energy of order $\bar{\epsilon}_{A\gamma}=18.31A_{56}^{-0.21}$MeV and a width of $\Delta \epsilon = 8\,$MeV), and a roughly constant tail, with a factor of 8 times lower cross section, at higher energies \citep{PSB76,Murase08}. The cross-section at the resonance can be written as $\bar{\sigma}_{A\gamma}\approx 8\times 10^{-26}\,A_{56}\,{\rm cm}^{2}$.
The threshold energy is of order $(\epsilon_1+\epsilon_2)/2=\bar{\epsilon}_{A\gamma}=18.31A_{56}^{-0.21}$MeV and $\epsilon_2 -\epsilon_1 = \Delta \epsilon$ \citep{PSB76,Karakula93,Murase08}. 
The corresponding threshold Lorentz factor for particles thus reads
\begin{equation}\label{eq:gamma_thres}
\gamma_{A, {\rm thres}} = \frac{\bar{\epsilon}_{A\gamma}}{\epsilon_{\gamma,{\rm th}}}\sim 2.1\times 10^5\,A_{56}^{-0.21} \,T_6^{-1}\ .
\end{equation}
For typical parameters appropriate for a highly magnetised fast spinning pulsar, this threshold energy is lower than the maximum energy $\gamma_{\rm max}$ reached by particles accelerated in the gap (Eq.~\ref{eq:gmax}), implying that most particles are in a condition to experience photo-hadronic processes. \\

The photo-hadronic interaction time on the neutron star thermal background can be expressed for a nucleus of mass number $A$, with Lorentz factor $\gamma_A$ and velocity $\beta_A$, as
\begin{eqnarray}
t_A = \left[c\int_{\epsilon_{\rm min}}^\infty {\rm d} \epsilon \,\frac{ {\rm d}n(\epsilon,T)}{ {\rm d} \epsilon}   \int _{\mu_{\rm M}}^1  \sigma_{A\gamma}(\epsilon')\,(1-\beta_A\mu)\, {\rm d} \mu \right]^{-1}
\end{eqnarray}
with $\epsilon' =  \gamma_A(1-\beta_A\mu)\epsilon$ (primed quantities are in the nucleus rest frame).  Here $\mu = \cos \theta$, with $\theta$ the angle between the direction of the cosmic ray and the photon (in the lab frame). The maximum angle from which a thermal photon produced at the surface of the star can hit a nucleus at a distance $ct$ from the surface is then $\mu_{\rm M}(t) = [1-R_\star^2/(R_\star+ct)^2]^{1/2}$. The anisotropic orientation of the photon background field radiated from the stellar surface is taken into account by a suitable choice of the integration limits in the integral over $\mu$. The photon number density ${\rm d}n(\epsilon,T)/{\rm d}\epsilon \approx U_{\rm th}/\epsilon_{\gamma, \rm th} \delta(\epsilon-\epsilon_{\gamma, \rm th})$ is calculated at the surface of the star.

Since we will mostly be concerned with non resonant interactions, the relevant interaction timescale can be approximated by
\begin{equation}
t_A = t_A^*\, \chi_A(t) \ ,
\end{equation}
where
\begin{eqnarray}
\label{eq:ta*}
t_{A}^* &=&  \left(c\,\frac{\bar{\sigma}_{A\gamma}}{8}\frac{U_{\rm th}}{\epsilon_{\gamma, {\rm th}}}\right)^{-1}\sim 6\times 10^{-5}\,{\rm s} \, A_{56}^{-1}\,T_6^{-3}
\end{eqnarray}
and 
\begin{equation}
\label{eq:chi}
\chi_A(t)^{-1} =  \left\{
\begin{array}{ll}
8 \int _{\mu_{\rm M}(t)}^1 \,(1-\beta_A\mu)\, {\rm d} \mu& \quad \mbox{if } \epsilon_1 \le \gamma_A(1-\beta_A\mu)\epsilon \le \epsilon_2 \\
\int _{\mu_{\rm M}(t)}^1 \,(1-\beta_A\mu)\, {\rm d} \mu&\quad \mbox{if }  \gamma_A(1-\beta_A\mu)\epsilon > \epsilon_2 \ .
\end{array}
\right.
\end{equation}

We define the instantaneous opacity of the interaction as $\tau_A = t/t_A$. For the sake of illustration, at the stellar surface, where the radiation field can be considered as roughly isotropic ($\chi_A(t=0) \sim 1$ for interactions out of the Giant Dipole resonance, second line in Eq.~\ref{eq:chi}), this quantity reads $\tau_{A}^* =  {R_{\star}}/({ct_{A}^*})\sim 0.5\,R_{\star,6}  A_{56}\,T_6^3$.

For photo-pion production by nucleons, the same scheme as for the Giant Dipole resonance can be adopted with $(\epsilon_1+\epsilon_2)/2=0.3\,$GeV, $\epsilon_2 -\epsilon_1 = 0.2\,$GeV, and $\bar{\sigma}_{p\gamma}=4\times 10^{-28}\,{\rm cm}^2$~\cite{Waxman97}. The much smaller cross-section compared to photo disintegration makes this process irrelevant in all cases, as we will see in the following section.

\begin{figure}[!h]
\begin{center}
\epsfig{width=0.45\textwidth,file=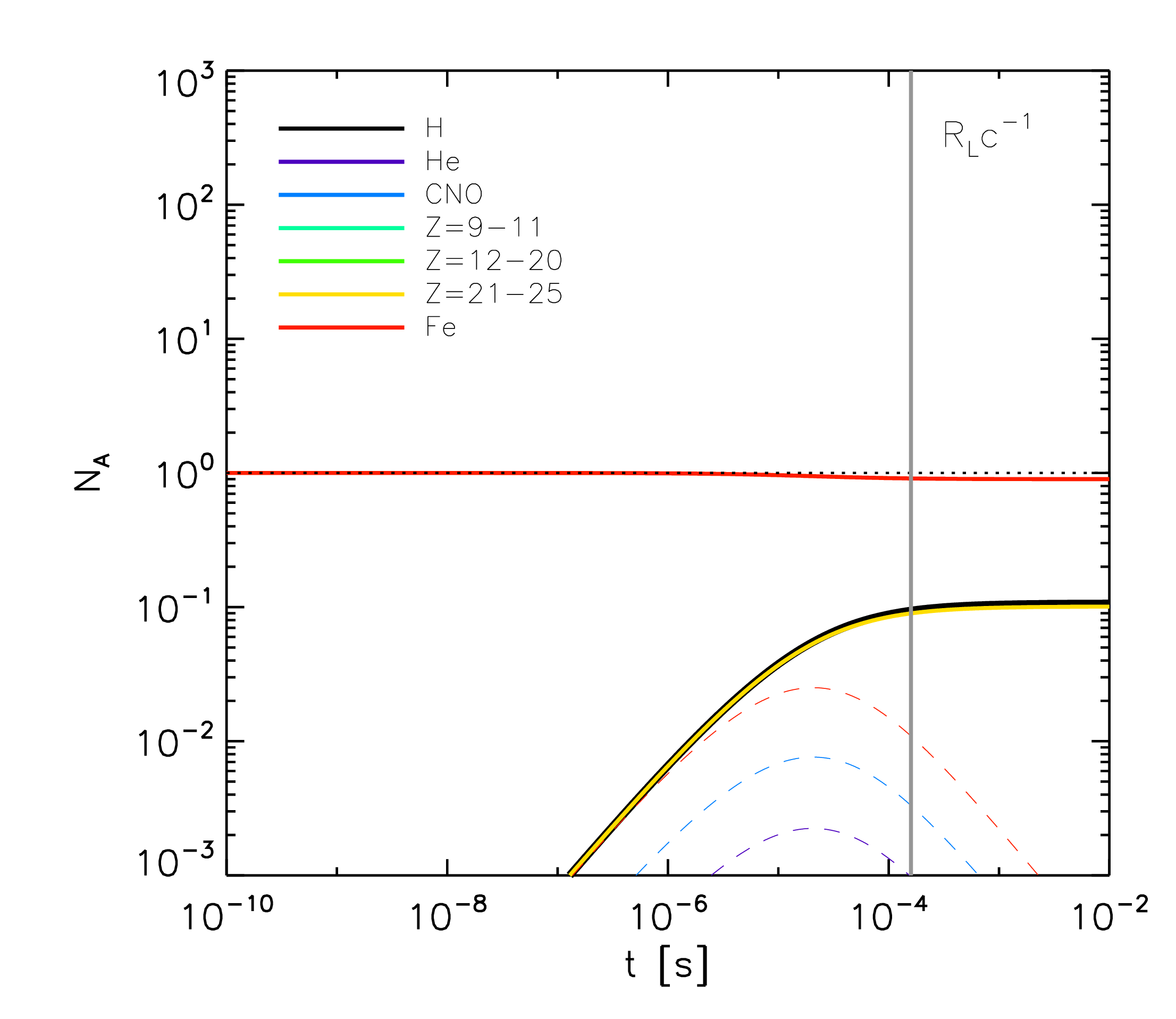} 
\epsfig{width=0.45\textwidth, file=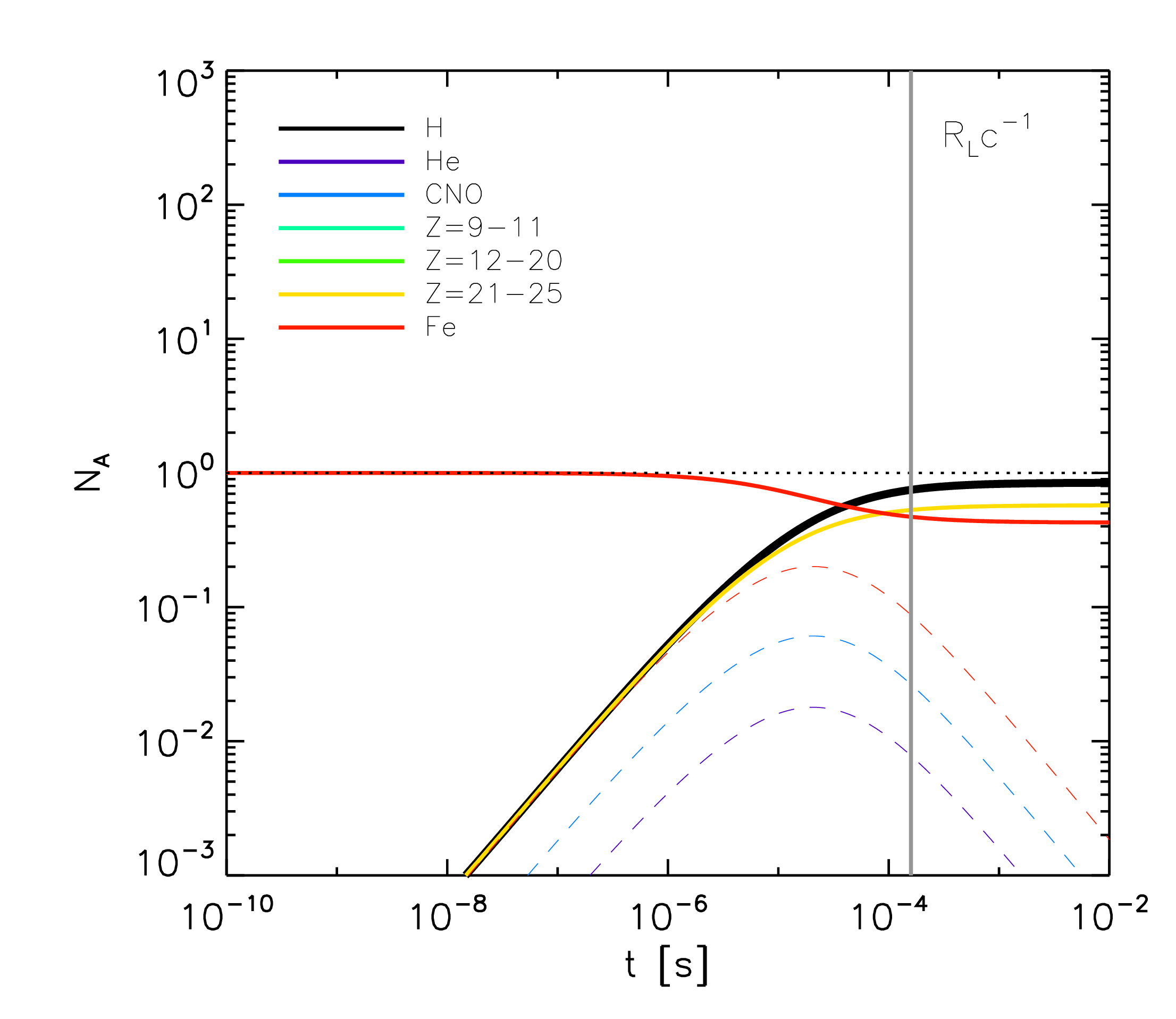} 
\epsfig{width=0.45\textwidth, file=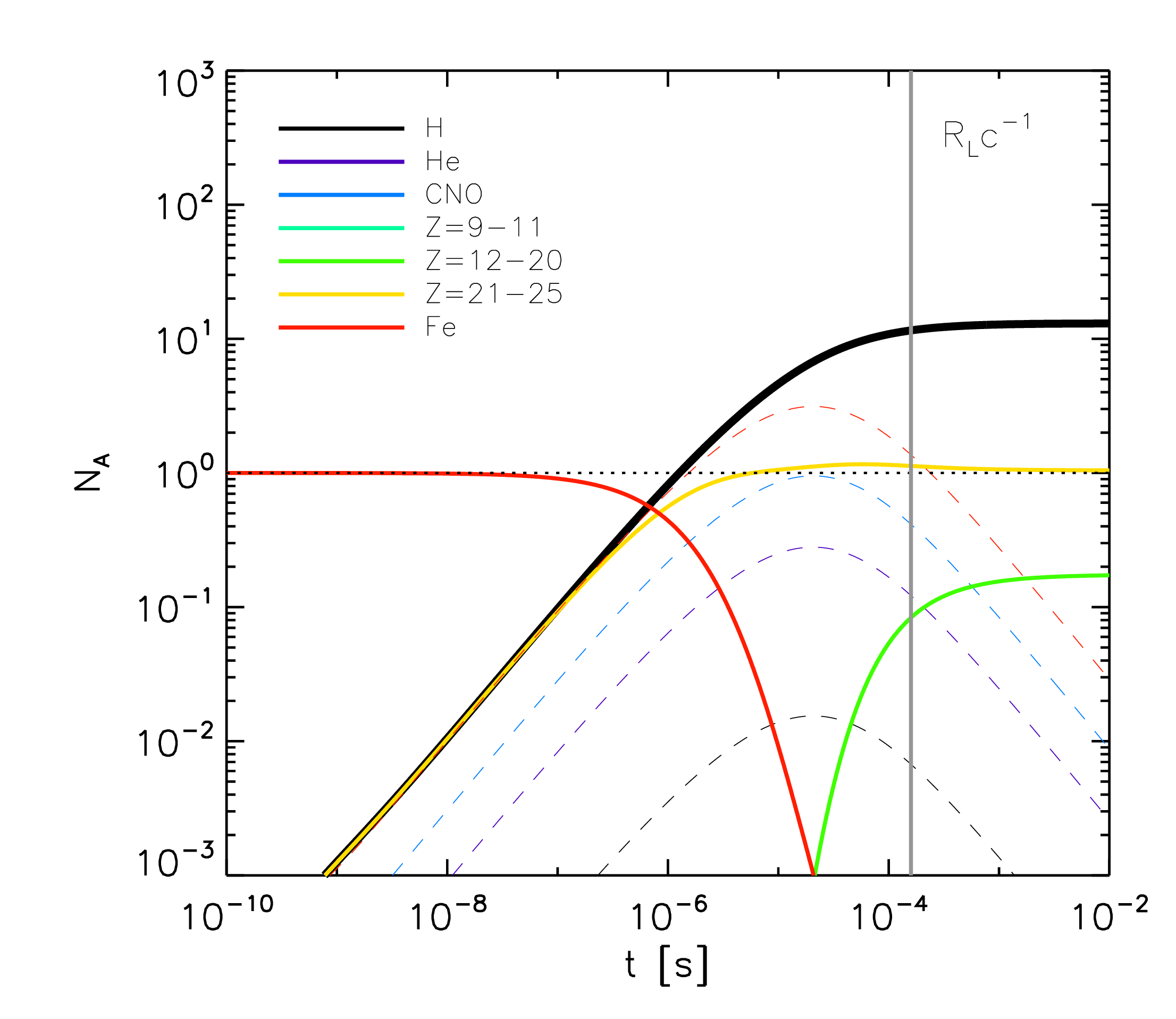} 
\epsfig{width=0.45\textwidth, file=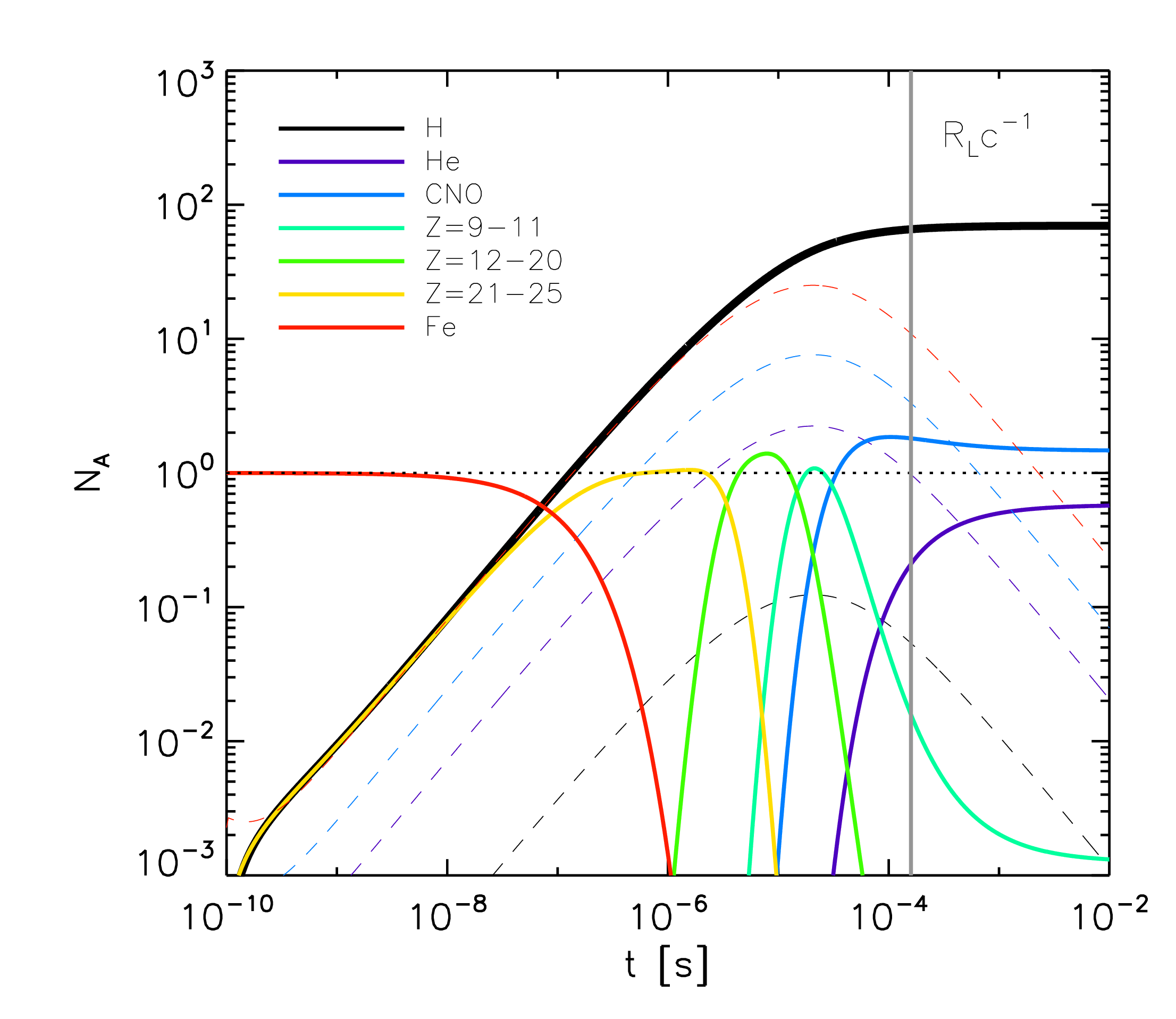} 
\caption{Composition of UHECR nuclei after photo-disintegration in the thermal radiation field of the star (solid lines). Extraction of pure iron at $t=0$ is assumed. Eqs.~\ref{eq:nfe} and \ref{eq:nat} are solved for $\gamma(t)$ determined by Eq.~\ref{eq:dgdt} (with $\xi=1$) up to the value $\gamma_{\rm max}$ set by Eq.~\ref{eq:gmax}.
The fractions of different species are represented by solid lines, with colours as specified in the legenda of each panel, while dashed lines refer to the opacities {\kumiko $\tau_{A}=t/t_{A}$}. From left to right, top to bottom, temperatures are $T=[1,2,5,10]\times 10^6\,$K. The vertical line indicates the time at which particles reach the light cylinder $R_{\rm L}c^{-1}$ and the dotted horizontal line indicates $N_A = 1$.
\label{fig:NA_ani}}
\end{center}
\end{figure}

\subsection{Primary and secondary nuclei ratios}

In what follows, we calculate the number of primary and secondary nuclei that become part of the pulsar wind, after particles have been extracted from the surface of the star, and have suffered photo-disintegration. We assume for simplicity that nuclei lose one nucleon in each interaction. We assume that pure iron is injected at time $t=0$ at the surface of the star. 

The evolution of the iron abundance then follows:
\begin{eqnarray}
N_{\rm Fe} &=& N_{\rm Fe}(t=0)\,\exp\left[-\frac{1}{t^*_{\rm Fe}}\int_{t=0}^{t} \frac{{\rm d}t'}{\chi(t')}\right]\ .
\label{eq:nfe}
\end{eqnarray}
For the other species with mass number $A>1$, one can write
\begin{equation}
\frac{{\rm d}N_A}{{\rm d}t} + \frac{N_A}{t_A} = \frac{N_{A+1}}{t_{A+1}}\ ,
\end{equation}
which, assuming $N_A(t=0) = 0$ for $A<56$, leads to:
\begin{equation}
N_A = \exp\left[-\frac{1}{t_A^*}\int_0^t\frac{{\rm d}t'}{\chi(t')}\right]\int_0^t \frac{N_{A+1}}{t_{A+1}}\exp\left[\frac{1}{t_A^*}\int_0^{t'}\frac{{\rm d}t''}{\chi(t'')}\right]\, {\rm d} t' \ .
\label{eq:nat}
\end{equation}
Figure~\ref{fig:NA_ani} shows the evolution of $N_A$ calculated via this equation with $t_{A\star}$ and $\chi$ defined as in Eqs.~\ref{eq:ta*} and \ref{eq:chi}. For surface temperatures exceeding $10^6$K, photo-disintegration starts to be effective and with increasing temperature an increasingly large fraction of the initial iron is transformed into lighter nuclei. When the star temperature is of order $10^7$K, no iron is left at the light cylinder, nor elements with $Z>10$. In this case the composition is mostly protons and $\sim 10\%$ CNO. The trend shown by the opacity of the different nuclei is readily understood: initially it simply increases with time, but then it saturates and starts dropping due to the decreasing photon field density. {The maximum opacity is reached at a time which is independent of nuclear species and stellar temperature, as can be readily understood from Eqs.~\ref{eq:ta*}$-$\ref{eq:chi}.}

One thing to notice is that, as we already anticipated at the end of the previous section, the photo-pion process is always irrelevant, as shown by the dashed black line representing the corresponding opacity in the different panels.

These photo-hadronic interaction calculations are done under rough approximations. We have considered a cross-section composed of a delta function (to mimic the Giant Dipole Resonance) and a plateau at high energies (for higher-energy processes), and assumed that only one nucleon is liberated in each interaction. In practice, at the energies that we consider, we barely fall in the Giant Dipole Resonance regime, but in the ``plateau'', where Quasi-Deuteron and Baryonic Resonance processes dominate. The products stripped from the nuclei via these interactions are likely not single nucleons, but deuterons or alpha particles. Fragmentation of the nuclei in several lighter nuclei is also possible. This constitutes a limit of our calculation. A more thorough numerical calculation with tabulated processes would enable to better assess the resulting composition. A better treatment, however, would not change our main result: heavy nuclei would still survive for temperatures $T\lesssim 10^7\,$K.

\section{Spectra injected in the wind}\label{section:spectra}

Particles accelerated in the gap are injected in the pulsar wind, and reach higher energies, with fluxes that we estimate below. They subsequently reach the nebula region and the supernova ejecta, where they encounter other radiative and baryonic fields. As mentioned in the introduction, the effects of these backgrounds were calculated extensively in \cite{FKO12,FKO13,LKP13}. More precisely, Ref.~\cite{FKO13} estimated that the typical composition that has to be accelerated in the wind region before being altered by the supernova ejecta and the propagation in the intergalactic medium was $\sim 50\%$ protons, $\sim 30\%$ CNO and $\sim 20\%$ Fe. These authors discuss that this result is rather robust (the composition weakly depends on the supernova parameters or on the narrow range of pulsar parameters allowed for successful UHECR acceleration). 

Neutron stars spin down and their rotational energy is channelled via their winds towards the outer medium. 
In the following we assume, as the observations suggest for the Crab pulsar, that the electromagnetic luminosity, $\dot E_{\rm p}=(8\pi B_\star^2 R_\star^6)/(3c^3P^4)$ \cite{Shapiro83}, is efficiently converted into kinetic luminosity, $\dot N \gamma_{\rm w} mc^2$. We write the latter as: $\dot N \gamma_{\rm w} mc^2\,\equiv\, 2\kappa m_e c^2\dot N_{\rm GJ} \gamma_{\rm w}\left(1+x_A\right)$, with $\dot N_{\rm GJ}=e^{-1}(\dot E_{\rm p}c)^{1/2}$ the Goldreich-Julian charge density \cite{Goldreich69}, and $x_{A}\,\equiv\, m_A/\left(2\kappa Zm_e\right)$, where $m_{A}=Am_p$ is the ion mass and $\kappa$ the pair multiplicity, namely the number of pairs produced for each electrons that leaves the star.

We can then write the wind Lorentz factor as \cite{LKP13}:
\begin{eqnarray}
\label{eq:gwind}
\gamma_{\rm w}(t)\,&\simeq&\,(1-\eta_B)\frac{\dot E_{\rm p}}{\dot N m c^2}\label{eq:gamma_wind}\nonumber\\
&\sim& 3\times 10^{9}\,(1-\eta_B)(1+x_{A})^{-1} (1+t/t_{\rm sd})^{-1} \kappa_4^{-1}P_{\rm i,-3}^{-2}B_{13}R_{\star,6}^3\quad .
\end{eqnarray}
Here, we have expressed the pulsar spin down luminosity $\dot E_{\rm p}(t)= \dot E_{\rm p0}/(1+t/t_{\rm  sd})^{2}$  (assuming a breaking index of 3), where the initial luminosity is $\dot E_{\rm p0}=E_{\rm  rot}/t_{\rm sd}\,\simeq\,6.4\times10^{44} P_{i,-3}^{-4}B_{\star,13}^2R_{\star,6}^6\,$erg/s, and the spin-down time is 
\begin{equation}
t_{\rm sd} = \frac{9Ic^3P_{\rm i}^2}{8\pi^2B^2R^6} \sim 3.1\times 10^{7}\,{\rm  s}\,I_{45}B_{13}^{-2}R_{\star,6}^{-6}P_{{\rm i},-3}^2\ . 
\end{equation}
$\eta_B$ represents the fraction of magnetic energy injected into the energy conversion region. Observations indicate that close to the pulsar wind nebula $\eta_B\ll 1$ \cite{Kirk09}.

Depending on the value of $\kappa$, that can range between $10-10^8$ in theory (a highly debated quantity, see e.g., \cite{Kirk09}), the energy conversion can be efficient enough to enable iron to reach energies at neutron-star birth ($t=0$)
\begin{equation}
E_0 \sim 2.3\times10^{20}\,{\rm eV} A_{56}\, \eta\, \kappa_4^{-1}P_{{\rm i},-3}^{-2}B_{13}R_{\star,6}^3 \ ,
\end{equation}
assuming $x_{A}< 1$ (we have, e.g., $x_{\rm Fe}\sim 0.2$ for $\kappa=10^4$) and $\eta_B\ll 1$. To take into account all the uncertainties on the luminosity conversion efficiency, we have introduced the efficiency factor $\eta\le 1$.\\

As the neutron-star spins down, cosmic rays flowing with the wind will have an energy
\begin{eqnarray}
E_{\rm CR} (t) &=& E_0\,\left(1+t/t_{\rm sd}\right)^{-1} \\ 
&\sim& 1.2\times10^{20}\,{\rm eV} \,\eta A_{56}\kappa_4I_{45}B_{13}^{-1}R_{\star,6}^{-3} \,{t_{7.5}}^{-1}\,\quad {\rm for}\ \  t>t_{\rm sd}\nonumber .
\label{eq:Ecrt}
\end{eqnarray}
Channelling the Goldreich-Julian charge density into particles and taking into account the neutron-star spin down rate (neglecting gravitational wave losses), one can write the cosmic-ray injection flux as:
\begin{equation}\label{eq:spectrum_arons}
\frac{{\rm d} N_{\rm CR}}{{\rm d} E} =\int_0^\infty {\rm d}t \dot N_{\rm GJ} (t) \delta \left(E-E_{\rm CR}(t)\right)=
\frac{\dot N_{\rm GJ}(0) t_{\rm sd}}{E}\, .
\end{equation}

\begin{figure}[!h]
\begin{center}
\includegraphics[width=0.45\textwidth]{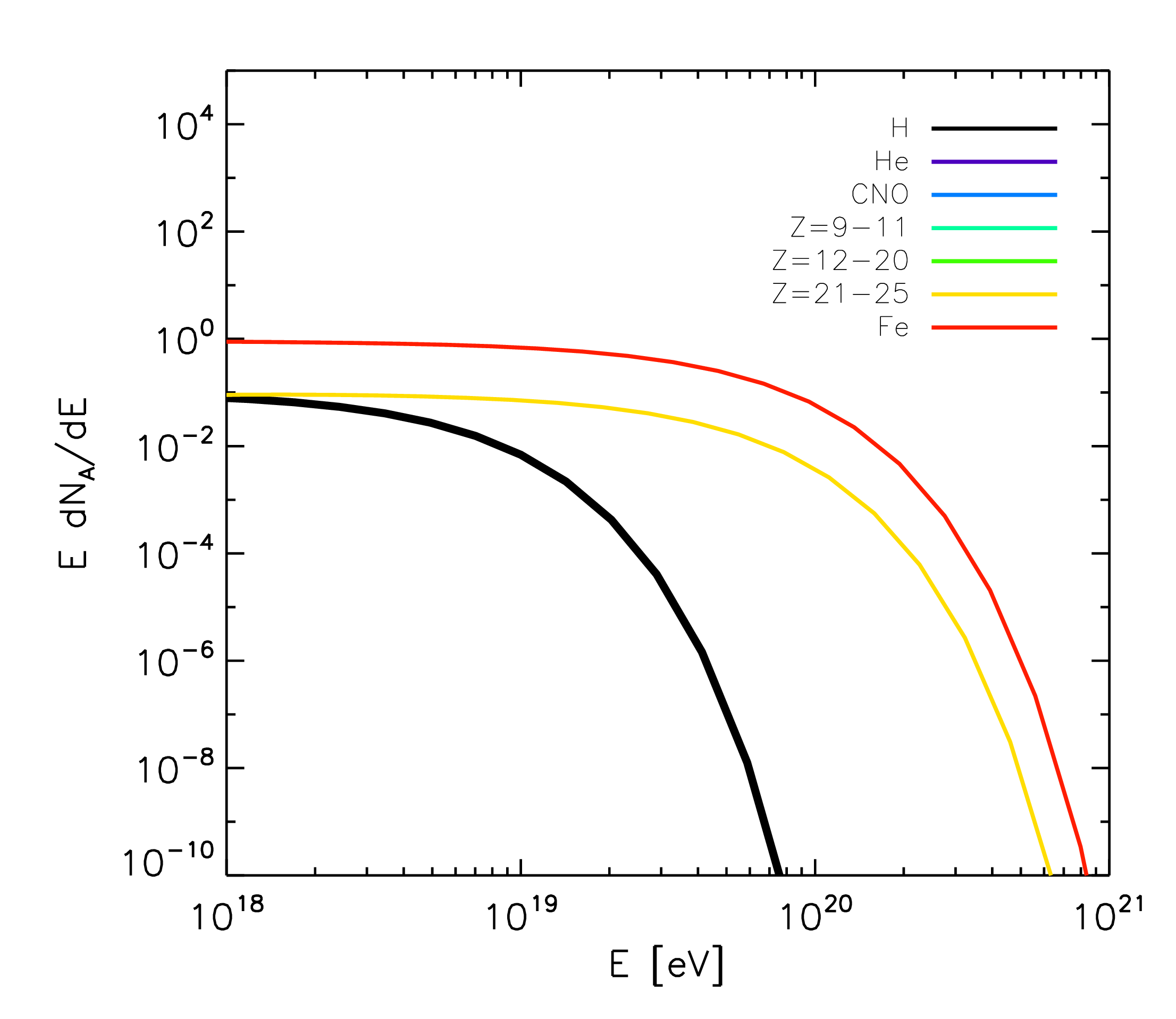} 
\includegraphics[width=0.45\textwidth]{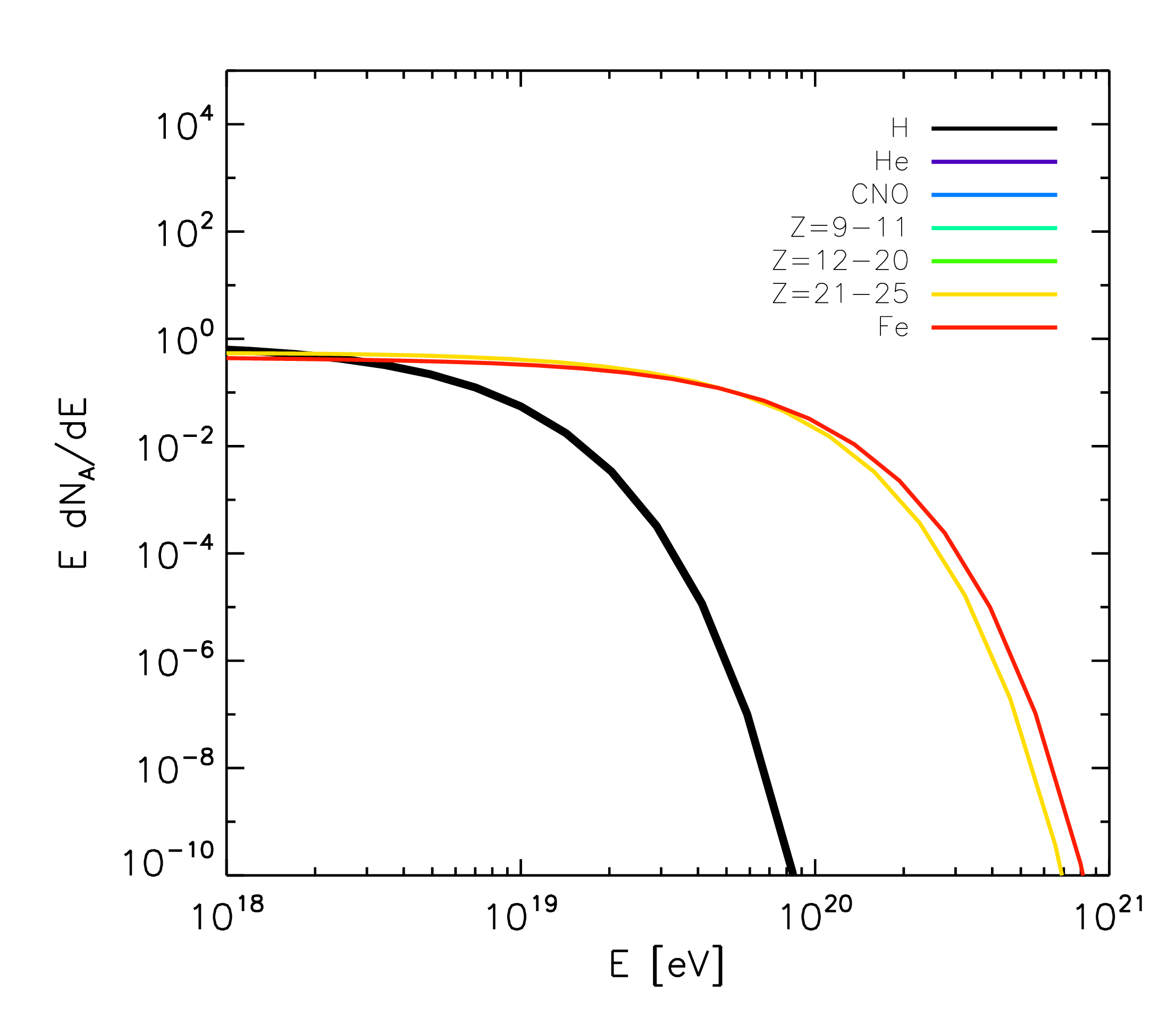} 
\includegraphics[width=0.45\textwidth]{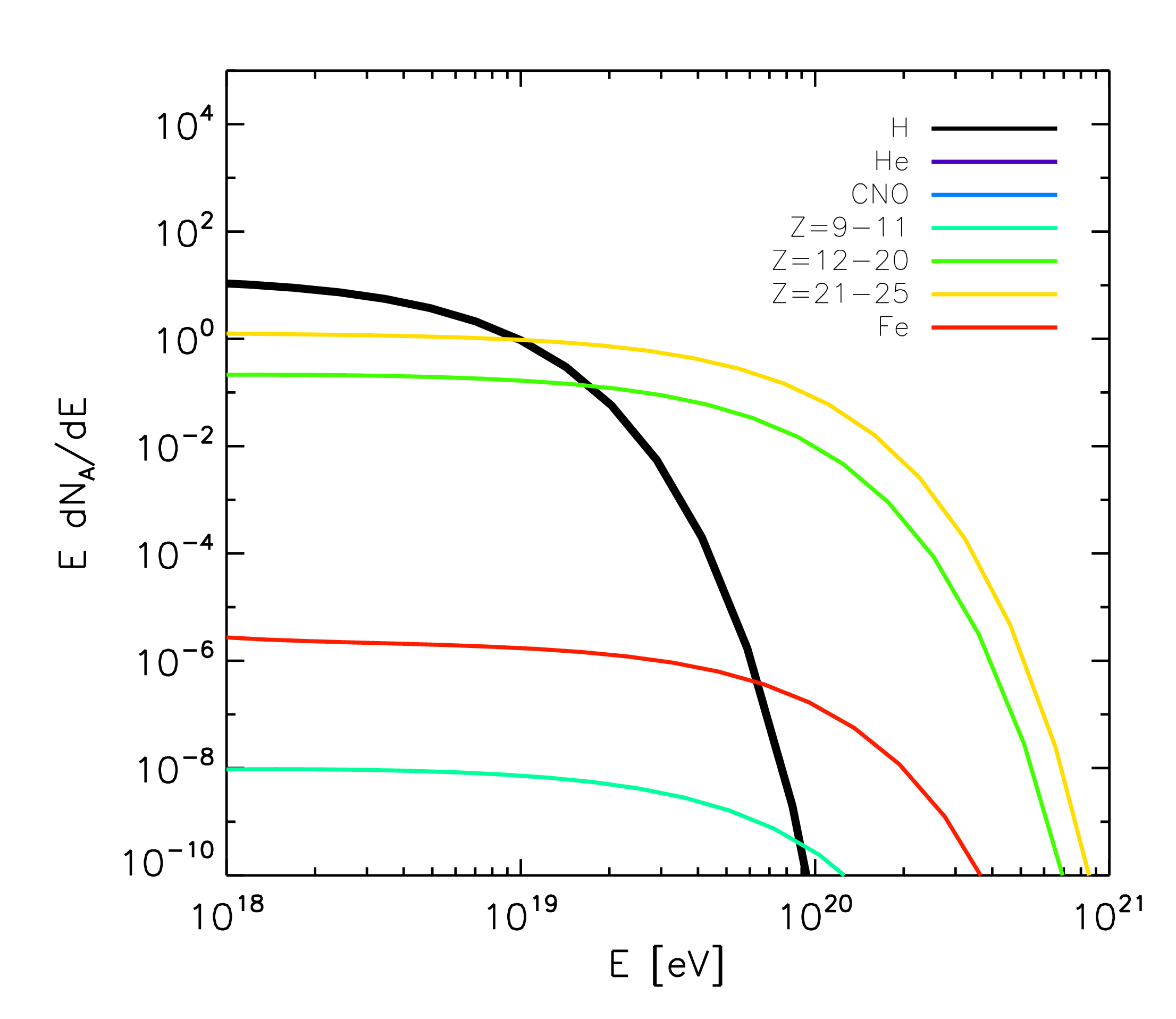} 
\includegraphics[width=0.45\textwidth]{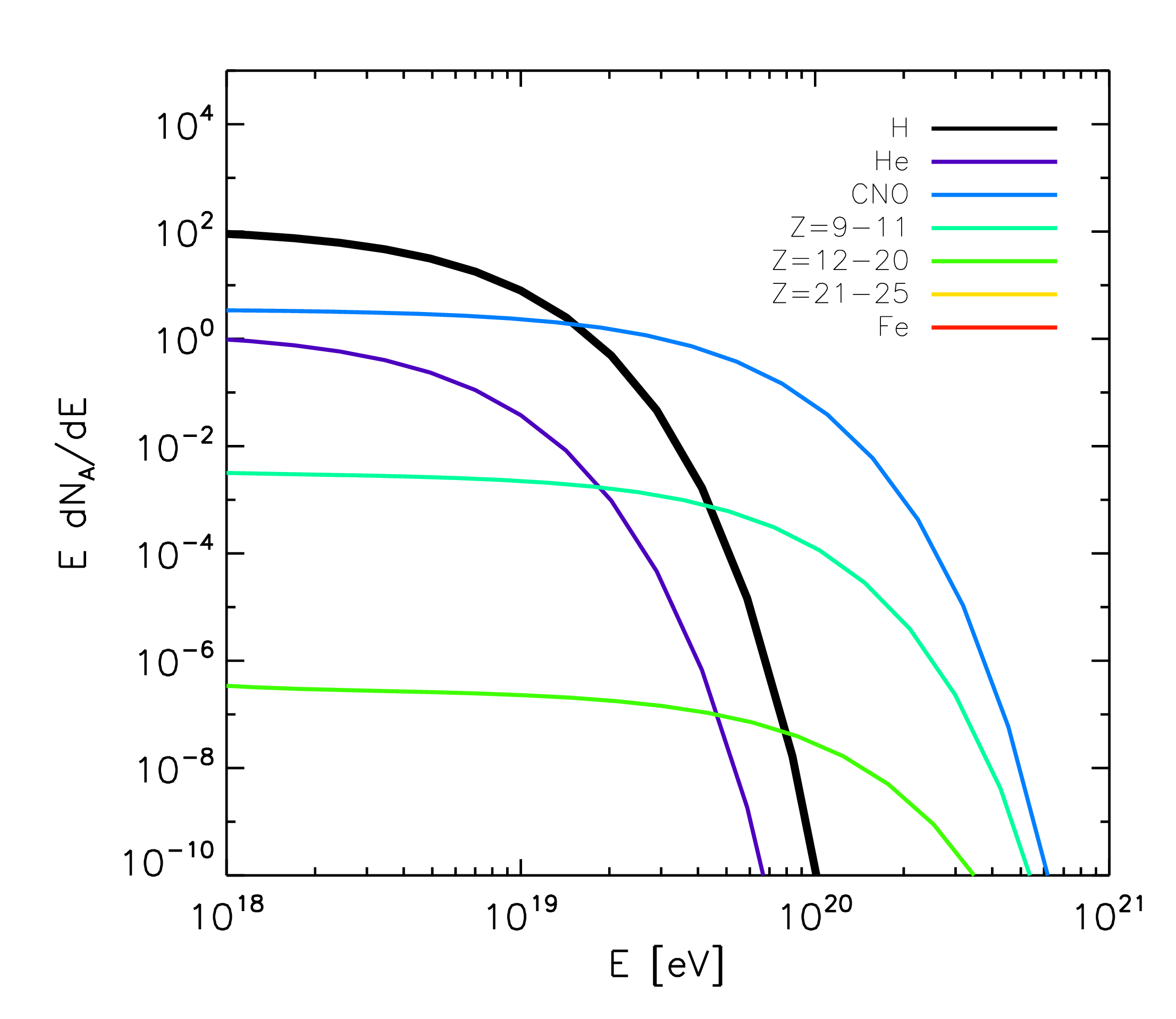} 
\caption{Cosmic-ray spectra $E{\rm d}N_A/{\rm d} E$ and composition estimated at time {$t_{\rm w}$ after injection with $t_{\rm w}=100\times R_{\rm L}/c$ and an injection time $<100\,$yr}. This limit on $t_{\rm inj}$, which reflects in the iron spectrum to be cut off at $\approx 10^{18}$ eV, was chosen because at later times the unknowns are such as to make our calculation unreliable (see text for details). The results refer to a case in which only iron is extracted from the star surface and photo disintegration is taken into account as described in Section~\ref{section:photodis}. From left to right, top to bottom, temperatures are $T=[1,2,5,10]\times 10^6\,$K.}
\label{fig:spectra_ani}
\end{center}
\end{figure}

In order to compute the spectrum and composition emerging from the star, we can use the results of the previous section and Eq.~\ref{eq:spectrum_arons} as long as a number of conditions are satisfied:
\begin{itemize}
\item the star surface temperature should not decrease appreciably below the value initially assumed
\item the electric field at the star surface, which decreases with time as $P^{-1}$ must stay larger than the value required for extraction of iron nuclei (which is also changing and likely increasing with time, due to the decreasing temperature)
\item the acceleration time (Eq.~\ref{eq:tacc}) needed to reach $\gamma_{\rm th}$, which has a strong dependence on the pulsar period, $\propto P^3$, should still be short enough so that $\gamma_{\rm th}$ is reached before the time corresponding to the peak in the opacity (which always occurs at $t\approx 10^{-2} R_{\rm L}(0)/c$, as one can see from Fig.~\ref{fig:NA_ani}).
\end{itemize}

While the evolution of the star surface temperature is not very well constrained, it seems reasonable to assume that this stays constant for the first few $\times 100\,$yr. If this is the case, then all other conditions are satisfied for $t<100\,$yr, and we decided to limit ourselves to this time. This causes us to be able to reliably compute the spectrum of iron only down to an energy $E\approx 10^{18}$ eV (see Eq.~\ref{eq:Ecrt}). This does not mean that the pulsar model predicts a low energy cut-off for the spectrum of iron at around this energy, with the UHECR composition predicted by this model becoming progressively lighter at lower energy. A low energy cut-off for iron is expected, due to the fact that condition 2 will be violated at some point, but currently we are simply not able to assess this matter, due to the large number of unknowns.
 
Figure \ref{fig:spectra_ani} shows the spectra of the primary and secondary nuclei produced in the wind at times {$t_{\rm w}=100\times R_{\rm L}/c$ after injection, for a primary iron nucleus injection at $t=t_{\rm inj}$ with $t_{\rm inj}<100\,$yr}. The time $t_{\rm w}$ is chosen to be much longer than the photodisintegration loss time scale, to ensure that these spectra reflect the composition that will be injected in the wind and leave the star with the wind Lorentz factor. In our calculations we use the output obtained in Section~\ref{section:photodis} and Eq.~(\ref{eq:spectrum_arons}) normalized to the total number of injected nuclei:
\begin{equation}
E\frac{{\rm d} N_{\rm A}}{{\rm d} E} = \frac{N_A}{N_{\rm Fe}(t=0)} \exp\left(-{\frac{\gamma}{\gamma_{\rm w}}}\right)\, ,
\end{equation}
where $\gamma_{\rm w}$ is defined in Eq.~(\ref{eq:gamma_wind}) with $x_A<1$ for all nuclei.

\begin{figure}[h]
\begin{center}
\includegraphics[width=0.6\textwidth]{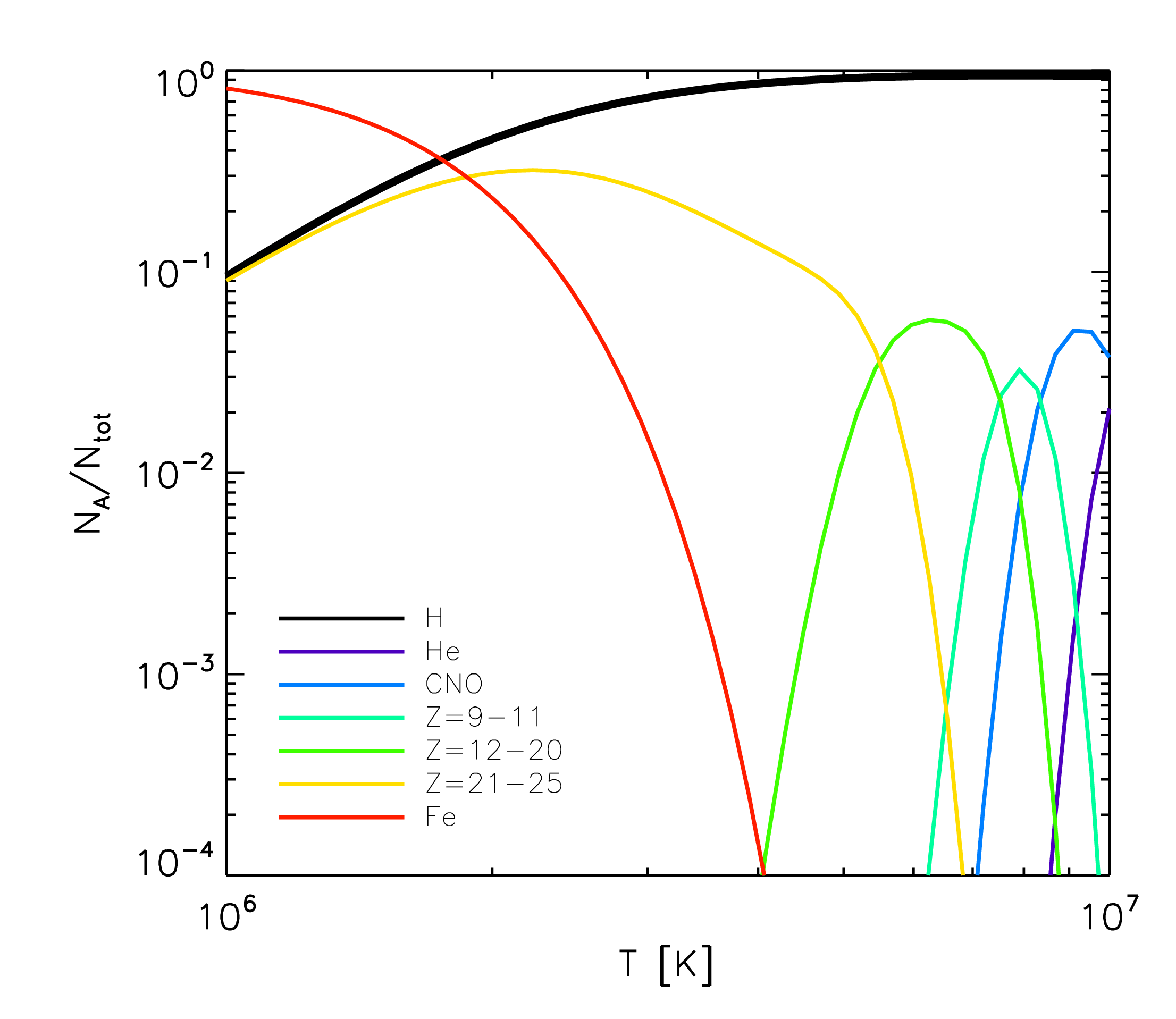} 
\caption{Fraction of each nuclear species injected in the wind at time $t_{\rm w}$ for energies $\gamma<\gamma_{A,\rm max}$, as a function of the neutron star surface temperature $T$, assuming initial pure iron injection. }\label{fig:Afrac}
\end{center}
\end{figure}

Figure~\ref{fig:Afrac} gives the fraction of each nuclear species as a function of the neutron star surface temperature $T$, assuming again that only iron is directly extracted from the stellar surface. This figure demonstrates that a significant fraction $\gtrsim 10\%$ of heavy nuclei survive the photo-disintegration process on the neutron star thermal background, for temperatures $T\lesssim 5\times 10^6\,$K. For higher temperatures, lighter nuclei are injected together with protons at a rate of $\sim 5\%$. It is interesting to note that this is a reasonable range of temperatures based on our knowledge of neutron star surface, and it would also provide a composition compatible with that required by Ref.~\cite{FKO13} to fit the UHECR data.

\section{Discussion and conclusions}\label{section:discussion}

The sources of UHECRs are still to be found. The observational situation has changed quite a bit in the last few years, leading to new challenges. The main new pieces of information that have arisen from Auger data can be summarised as follows: 1) The mass composition, as measured from the the elongation rate $X_{\rm max}(E)$ and its fluctuations, drifts from a lighter one at $E\sim 10^{18}$ eV to a heavier one at the highest energies \cite{Auger_icrc13,TA_auger_icrc13,FKO13}; 2) The injection spectrum required to fit the fluxes and the mass composition at the same time seems to be harder than the typical $\sim E^{-2}$ inferred for most astrophysical sources \cite{Aloisio14}; 3) If taken to scale as the charge $Z$ of the nucleus, the maximum energy is bound to be $\sim 5\times 10^{18}\,Z$\,eV \cite{Aloisio14}, thereby relaxing the tight constraints on the maximum energy in case of pure proton composition.

All of these three requirements appear to be fulfilled if fast spinning very young pulsars are the sources of UHECRs: as recognized already in early work on this scenario \cite{Blasi00,Arons03}, the spectrum of particles accelerated in the pulsar magnetosphere is very hard, typically $\propto E^{-1}$. Since the particles that get energized are extracted from the the iron-rich surface of the star, it is likely that the initial composition of the accelerated particles is dominated by Fe nuclei~\cite{Blasi00}, although propagation in the ejecta of the parent supernova may change the abundances of lighter elements \cite{FKO13}.

Within the pulsar scenario for the origin of UHECRs, one fundamental question that has never been convincingly addressed, in the relevant parameter range, is that of Fe survival in the hostile environment provided by the close magnetosphere. This has represented the main goal of the present paper: we have calculated the effects of the immediate pulsar environment on heavy nuclei, focusing on the dominant background, represented by the thermal radiation field from the star. The electric field due to the fast rotation of the neutron star gives rise to the extraction of iron nuclei from the star surface. Even in the simple case of pure Fe composition of the stripped off material, we found that a mixed composition is likely to be finally produced, as a result of photo-disintegration processes in the magnetosphere of the pulsar. For reasonable assumptions on the star temperature ($T\sim {\rm few} \times 10^6\,$K) and stellar parameters, it is possible to have a relative abundance of the injected elements that is in qualitative agreement with that required to reproduce the mixed composition observed by Auger (see Ref.~\cite{FKO13}). {In particular, the intermediate mass composition suggested by the latest Auger data \cite{Auger_compo_2014a,Auger_compo_2014b}, can be easily accommodated for stellar surface temperatures around $\sim 5\times 10^{6}$ K.}

Our calculations also show that at times of order $\sim 100$ years, it becomes challenging for the induced electric field at the star surface to overcome the binding energy of iron nuclei in the lattice, so that a low energy cutoff (or suppression) is induced in the spectra of injected iron, typically in the EeV energy range. Lower energy particles can still be produced as a result of photo disintegration of iron nuclei in the magnetosphere. 

Once the nuclei have been stripped off the surface of the star, they get accelerated in the gap, assumed to be distributed on a region of size $R_{L}$ (radius of the light cylinder) from the stellar surface. While being accelerated, particles suffer photo-disintegration processes on the photon background emitted by the hot surface of the star, and lose energy to curvature radiation. Curvature energy would limit the Lorentz factor of the nuclei to $\sim 10^{8}$, which for fast spinning neutron stars is appreciably lower than the value reachable by using the full potential drop. However, the nuclei that reach the light cylinder region eventually end up in the wind of electron-positrons propagating outwards. If the multiplicity of electrons and positrons is $\lesssim 2m_{p}/m_{e}$, the maximum energy of iron nuclei (and of the secondary nuclei that are produced by photo disintegration) is in the range of interest for UHECRs. Notice that by construction the maximum energy of the secondary nuclei scale proportional to the mass number $A$ instead of the charge $Z$ of the nucleus.

\section*{Acknowledgements}
We thank Roberto Aloisio for fruitful discussions. KK thanks the Osservatorio Astrofisico di Arcetri, where part of this work was done, for its very kind hospitality. KK acknowledges financial support from Sorbonne Universit\'es, the Institut Lagrange de Paris and PNHE. The work of EA and PB was partially funded through Grant PRIN-INAF 2012.

\label{sec:acknowledgements}

\bibliography{photodis}
\end{document}